\documentclass[5p]{elsarticle}
\journal{Nuclear Physics B}
\biboptions{sort&compress} 
\usepackage{amsmath}
\usepackage{graphicx}
\usepackage{amssymb}
\usepackage{tabularx}

\usepackage[colorlinks=true, linkcolor=blue, citecolor=blue, urlcolor=blue]{hyperref}

\begin{document}

\begin{frontmatter}

\title{Enhanced extragalactic photon flux by PBH in a stimulated axion/ALP decay scenario}

\author[1]{Yadir Garnica\corref{cor1}}
\ead{ya.garnicagarzon@ugto.mx}

\author[1]{J. Barranco}
\ead{jbarranc@fisica.ugto.mx}

\author[1]{Luis A. Ure\~na-L\'opez}
\ead{lurena@fisica.ugto.mx}

\cortext[cor1]{Corresponding Author}

\address[1]{Departamento de F\'isica, DCI, Campus Le\'on, Universidad de Guanajuato, Le\'on, 37150, M\'exico}

\begin{abstract}

In this work, we show that stimulated decay of axions or axion-like particles in black hole superradiance is an efficient way to find and hunt primordial black holes. When de Broglie's wavelength of the axion/ALP is comparable or larger than the black hole horizon radius, a large population of such particles accumulates in the surroundings of the black hole. When these axions/ALPs couple to photons, the bosonic cloud decays into photon pairs, generating intense stimulated radiation emission that contributes to the X-ray, visible light, and radio wave background flux that can exceed current observational limits measured at Earth. If the masses are in the interval of $10^{-3}\mathrm{eV}<\mu <1\mathrm{eV}$, to be consistent with current observations of microwave background light, we found that the fraction of primordial black holes should be smaller than $f_{PBH} < 10^{-17}$ for PBHs with masses within $10^{-13} M_{\odot}<M_{BH}<10^{-7} M_\odot$.    
\end{abstract}


\end{frontmatter}

\section{Introduction}

The de Broglie kilometer-size wavelength of an ultralight boson with masses below the nano $\mathrm{eV}/c^2$ scale induces exciting phenomena around black holes (BHs). In particular, candidates for ultralight spin-zero DM particles such as fuzzy DM \cite{Hu:2000ke,Matos:2000ss, Matos:2000ng,Hui:2016ltb,Urena_review,matos2023short}, axion or ALPs \cite{Marsh:2015xka} can trigger a superradiant instability in rotating black holes \cite{Zeldovich,Zeldovich72,penrose,Starobinsky:1973aij,Starobinskii:1973hgd,Brito:2014wla,East:2017ovw}. Furthermore, while massive scalar fields can form quasi-bound states that survive for times longer than the age of the universe even around spherically symmetric non-rotating black holes, the presence of angular momentum in the rotating case turns these bound states into a powerful instability.
\cite{Barranco2011,Barranco:2012qs,Barranco:2013rua,Barranco:2017aes,Aguilar-Nieto:2022jio,Alcubierre:2024mtq}.
Both, the instability in rotating BH and the survival of the bosonic particles is regulated by the dimensionless ``gravitational fine-structure constant" 
\begin{equation}
\alpha_G=G \mu M_{BH}\,,\label{fine}
\end{equation}
where $G$ is the Newton's, gravitational constant, $\mu$ is the boson mass and $M_{BH}$ is the mass of the black hole. Here $\hbar=c=1$. As long as $\alpha_G\ll1$ a cloud of bosonic, ultralight particles can survive in the proximity of BHs. 

Additionally, axion/ALP can decay into two photons due to the anomaly term. Recent theoretical developments have shown that the decay of particles in superradiant instabilities can transform black holes into extremely bright lasers \cite{Rosa:2018Sti, Ikeda:2019}. Motivated by these mechanisms and subsequent studies exploring the observable photon fluxes from superradiant bosonic clouds \cite{Ferraz:2020, Branco:2023, Calza:2023}, in this work we establish constraints on the fraction of PBHs as dark matter. 
This method is particularly useful for constraining PBH in the mass window  around $10^{-13}M_\odot$  to
$10^{-7}M_\odot$. 
Our limits are complementary to PBH constraints obtained by microlensing \cite{Paczynski:1985jf,Griest:2013aaa,Niikura:2017zjd}, femtolensing \cite{Barnacka:2012bm}, bounds obtained by nondetection of cosmic rays produced by Hawking radiation \cite{Boudaud:2018hqb, Dasgupta:2019cae,Ballesteros:2019exr} or by capture of PBH on neutron stars \cite{Capela:2013yf}. The method consists in computing the number of photons produced by the decay of axions or axion-like particles in the cloud around a spinning PBH and comparing the resulting flux with the current limits on photons at visible, radio, and microwave wavelengths \cite{Overduin:2004dark}.
Before presenting our constraints, it is crucial to explicitly emphasize that the black hole lasing mechanism, which constitutes the core of this paper, relies on a specific set of theoretical assumptions regarding the nature of PBHs and ALPs:
\begin{itemize}
       \item \textbf{PBH Spin and Formation Epoch:} The efficiency of the superradiant lasing mechanism relies on PBHs possessing high spin parameters. Rather than assuming a single specific cosmological origin, our model relies on the well-documented premise that a highly spinning PBH population can naturally arise through multiple mechanisms. Even if born with negligible spin in a radiation-dominated era, PBHs can acquire substantial spin through binary mergers \cite{Rosa:2018Sti, Ferraz:2020} or evaporation-induced spin-up \cite{Calza:2023}. Alternatively, they can be born with near-extremal spins if formed during an early matter-dominated phase \cite{Harada:2017}. Our constraints are thus applicable to the fraction of PBHs that reach this high-spin regime through any of these valid channels.

    \item \textbf{Existence of ALPs in the relevant mass range:} The mechanism assumes the existence of ultralight scalar fields (ALPs) whose masses $\mu$ geometrically match the inverse radius of the PBHs to efficiently trigger the superradiant instability \cite{Arvanitaki:2010Str, brito:2020SR}.
    \item \textbf{Model-dependent particle interactions:} Even if superradiant clouds form, the lasing effect strongly depends on the ALP parameter space. Specifically, the axion decay constant ($f_a$) and the electromagnetic coupling coefficient ($C_{a \gamma \gamma}$) must allow stimulated decay into photons to overcome non-linear self-interactions and photon escape rates \cite{Rosa:2018Sti, Branco:2023, Calza:2023}.
   \item \textbf{Dark Matter Distribution:} For the derivation of the constraints, we adopt the standard simplification that the PBH population has a monochromatic mass distribution \cite{Branco:2023, Cheek:2022, DeLuca:2019}. This approach allows us to decouple the constraints from the highly model-dependent details of PBH formation mechanisms.
  \item \textbf{Quasiadiabatic approximation and evaporation limits:} We work within a quasiadiabatic framework, where the absence of external interactions yields a scalar cloud of mass $M_c$ that grows by extracting rotational energy from a Kerr PBH of mass $M_{PBH}$. In this isolated PBH-cloud system, the total mass and angular momentum are strictly conserved. Consequently, the maximal mass extracted by the scalar cloud is bounded; theoretical estimates in this scenario place the maximum mass ratio at $M_c/M_{PBH} \simeq 10.78\%$ \cite{Hui:2023}. Furthermore, to avoid the constraints imposed by Hawking evaporation, we restrict our analysis to PBH masses above the evaporation threshold. This ensures the black holes survive to the present day, which corresponds to the requirement $M_{PBH} \gtrsim 10^{14}\,\mathrm{kg}$ \cite{Branco:2023, Cheek:2022, Calza:2023} as shown in Fig.~\eqref{fig:ALP_parameter_space}.

\end{itemize}
The structure of this work, in order to obtain such constraints, is the following: in section \ref{sec:review} we discuss the lasing produced by the axion/ALP decay in PBH as proposed in \cite{Rosa:2018Sti}. In section \ref{sec:fluxes} the flux of photons is computed considering that a fraction of DM is composed of a monochromatic distribution of PBHs. Finally, in section \ref{sec:Conclusions}, we conclude by showing that this mechanism yields bounds on the PBH abundance that are significantly stronger than previous ones, reaching $f_{PBH}<10^{-17}$ for masses in the range $10^{-13} M_\odot<M_{PBH}<10^{-7}M_\odot$.

\section{Lasing by PBHs}
\label{sec:review}

Axions and ALP can interact with photons through the Lagrangian
\begin{equation}
\mathcal{L}_{int}=-\frac{\alpha_{EM}C_{a\gamma\gamma}}{8 \pi f_a}\phi F_{\mu\nu}\tilde{F}^{\mu\nu} \, , \label{eq:Linteraction}
\end{equation}

The Lagrangian in Eq.~\eqref{eq:Linteraction} defines the coupling between the QCD-axion/ALP and photons. Based on this interaction, the authors in~\cite{Rosa:2018Sti} investigate the formation of bosonic particle overdensities around spinning black holes.
Using an initial constant distribution of bosons and a uniform photon occupation distribution, it was found that the coupled axion-photon system evolves according to the set of equations~\cite{Kephart:1986vc, Kephart:1994uy, Rosa:2018Sti}
\begin{subequations}
\label{eq:SDOE_NAG}
\begin{eqnarray}
\frac{dN_{a}}{dt} &=& (\Gamma_{s}-\Gamma_a) N_{a} - \Gamma_aN_\gamma \left[ AN_{a} - B_{1} N_{\gamma}\right] \,, \label{eq:SDOE_NAG1} \\
\frac{dN_{\gamma}}{dt} &=& 2\Gamma_{a}N_a+N_{\gamma}\left[2\Gamma_aAN_{a} -\Gamma_e-B N_{\gamma} \right] \, , \label{eq:SDOE_NAG2}
\end{eqnarray}
\end{subequations}
which includes a surface-loss of photons that contribute with an extra term on the right-hand side that is proportional to the photon cluster-crossing time, generating a set of photons escaping from the axion/ALP cloud at a rate 
\begin{equation}
\Gamma_{e} \equiv c/(\sqrt{5} r_{0}) = 1.5 \times 10^{15} \alpha_G \tilde{\mu} \, \mathrm{s}^{-1} \, ,
\end{equation}
where $r_0$ corresponds to the Bohr radius in the hydrogen-like image
\begin{equation}
    r_0=\frac{\hbar}{\mu c \alpha}\, ,
\end{equation}
and the $\sqrt{5}$ factor results from the selection of the leading emission level, which corresponds to the minimal instability time scale. The ALP decay rate~\cite{Bauer:2017Coll, Bauer:2019ax} given by
\begin{equation}
\Gamma_a=\frac{\alpha_{EM}^2 \left|C_{a\gamma\gamma}\right|^2 \mu^3}{256\pi^3 f_a^2}\, . \label{eq:decayR_ALP}
\end{equation}

The numerical coefficients are explicitly given by $A=8\alpha_G^2/25$, $B_1=2\alpha_G^4/75$ and $B=2\alpha_G^3(\alpha_G+6)/75$. The most important parameter in Eqs. \eqref{eq:SDOE_NAG}  is $\Gamma_s$, which measures the rate of accretion of the bosonic field by the black hole.  It is given by
\begin{subequations}
\begin{equation}
\Gamma_s=\alpha_G^{4l+4}\mu \left(\frac{a m}{M_{BH}}-2\mu r_+\right) C_{nlm}\, ,
\label{eq:Gamma_s}
\end{equation}
with $C_{nlm}$ given by \cite{Starobinskii:1973hgd,Starobinsky:1973aij,Detweiler:1980uk}
\begin{align}
    C_{nlm} &= \frac{2^{4l+2}(2l+n+1)!}{(l+n+1)^{2l+4}n!}\left(\frac{l!}{(2l)!(2l+1)!}\right)^2 \nonumber \\
     \times& \prod_{k=1}^l \left[k^2\left(1-\frac{a^2}{M_{BH}^2}\right)+\left(2r_{+}\mu-\frac{a m}{M_{BH}}\right)^2\right]\, ,
     \label{eq:cnlm}
\end{align} 
\end{subequations}
where,  $l$ and $m$ are the orbital and magnetic quantum numbers, respectively. The integer $n \ge 0$ denotes the radial quantum number, which counts the nodes in the radial profile of the scalar field $\phi$. The principal quantum number is defined as $\tilde{n} = n + l + 1$ and determines the energy spectrum of the bound states. The fundamental mode, which represents the leading emission level, corresponds to $n=0$ and $l=m=1$ (yielding $\tilde{n}=2$).  Here, $a$ is the black hole spin and $r_{+} = M_{BH} (1 + \sqrt{1 - (a/M_{BH})^2})$ is the event horizon radius.

If $m>0$, $\Gamma_s$ could be positive and in this case, any initial axion/ALP field grows exponentially. That is the so-called superradiant behavior.
The fastest growing mode for $\Gamma_s$ is given for $l=1, m=1,n=0$, and for this case
$C_{nlm}=1/24$ and consequently 
\begin{equation}
\Gamma_s = \frac{1}{24}\left(\frac{a}{M_{BH}} -2\mu r_{+}\right)\alpha_G^{8}\mu \, . \label{gamma_s}
\end{equation}
Now it is clear why if $\alpha_G\ll1$ then $\Gamma_s\ll1$: the black hole cannot accrete the bosonic field. 

The early time behavior of the coupled system~\eqref{eq:SDOE_NAG} can be found by neglecting the second-order terms $\mathcal{O}(N^2)$ in particle numbers and taking the initial conditions $N_a(0)=1$ and $N_\gamma (0)=0$ (one boson and no photons). The linear version of the system~\eqref{eq:SDOE_NAG} has a decaying mode of the form $e^{-\Gamma_e \tau}$, which we also neglect, and then the early time attractor is explicitly given by 
\begin{align}
    N_a(t) &= e^{(\Gamma_s - \Gamma_a) t} \, ,\\
    N_\gamma(t) &= \frac{2\Gamma_a}{\Gamma_s-\Gamma_a+\Gamma_e}N_a(t)\, .
    \label{eq:linear-sols}
\end{align}
Hence, the exponential growth of bosons is possible as long as $\Gamma_s > \Gamma_a$, see also Eqs.~\eqref{eq:decayR_ALP} and~\eqref{gamma_s}.

Assuming as an initial condition for the axions that $N_a(0)=1$, 
the primary production of axions will be driven by superradiance. 
That is, $N_a=e^{\Gamma_{s} t}$. 
Neglecting terms proportional to $N_{\gamma}^{2}$, it is then possible 
to assume the following initial condition for the axions:
\begin{equation}
N_a(t)\sim N_a(0)-\frac{1}{2}N_{\gamma},
\end{equation}
where $N_a(0)\sim e^{\Gamma_{s} t}$. 
Substituting $N_a(t)$ into the photon evolution equation gives
\begin{align*}
\frac{dN_{\gamma}}{dt}
  = -\Gamma_{e}N_{\gamma}
  + 2\Gamma_a\!\left[
      N_a(0)
      + A N_a(0)N_{\gamma}
      - \frac{1}{2}N_{\gamma}
    \right].
\end{align*}
If $N_{\gamma}(0)=0$, the initial behavior of the photon number is given by 
$N_{\gamma}=2\Gamma_aN_a(0)t$. 
For intermediate times 
\begin{align*}
\frac{dN_{\gamma}}{dt}
  = 2\Gamma_a N_a(0)
  + \kappa N_{\gamma}
  \quad\Rightarrow\quad
  N_{\gamma}(t)
  = \frac{
      2\Gamma_a N_a(0)
      \left(e^{\kappa t}-1\right)
    }{\kappa}\, ,
\end{align*}
where $\kappa=2\Gamma_a A N_a(0)-\Gamma_a-\Gamma_{e}$. Taking into account that 
\begin{equation}
\frac{\Gamma_a }{\Gamma_{e}}
  = \frac{1/\tau_{\phi}}{c/r_{0}}
  \sim \frac{r_{0}}{c\tau_{\phi}} \ll 1
  \quad\Rightarrow\quad
  \kappa\approx 2\Gamma_a A N_a(0)-\Gamma_{e},
\end{equation}
an exponential growth starts on an approximate time scale $\tau_{las}\sim1/(2\Gamma_{a}AN_{a}|_0-\Gamma_{e})$, so $\tau_{las}>0$ implies the existence of a critical number of axions \cite{kephart:1987luminous, kephart:1995stimulated}.

It is possible to compute the critical values of the number of particles at late times. For the axion/ALP particles, such an asymptotic value for the bosons is $N_{a}^{c}=\Gamma_{e} /2A\Gamma_{a}$~\cite{Rosa:2018Sti}, or more explicitly for our case,
\begin{equation}
N_{a}^{c}= \frac{2.98\times10^{40}}{|C_{a\gamma\gamma}|^2}\left(\frac{f_a}{10^8\mbox{GeV}}\right)^2\left(\frac{\mu}{\mbox{eV}}\right)^{-3}\left(\frac{M_{BH}}{10^{-8}M_\odot}\right)^{-1} \, .
\end{equation}
It is in this stage that stimulated decay takes over and lasing begins, and then the number of photons exhibits damped oscillations about the asymptotic value $N_{\gamma}^c = \Gamma_{s}/A\Gamma_{a}$, which is also explicitly given by
\begin{eqnarray}
\label{eq:photoncritical}
N_{\gamma}^c &=& \frac{4.34\times 10^{52}(a^*-2G\mu r_+)}{|C_{a\gamma\gamma}|^2} \nonumber \\
&& \times \left(\frac{f_a}{10^8\mbox{GeV}}\right)^2\left(\frac{\mu}{\mbox{eV}}\right)^4\left(\frac{M_{BH}}{10^{-8}M_\odot}\right)^6 \, .
\end{eqnarray}
Note that Eq.~\eqref{eq:photoncritical} establishes that systems with smaller values of $C_{a\gamma\gamma}$ require a longer stabilization time compared to those with larger coupling constants. This shows that the photon production process due to axion-photon coupling avoids a formation of bosenova-type phenomena in a more efficient way.
Furthermore, from Eq.~\eqref{eq:photoncritical} it is found that there is a maximum number of photons for a given mass of the ALP particle. This maximum value is obtained if the BH mass is such that 
\begin{equation}
M_{BH} =\frac{3}{7}\frac{a^*}{G\mu(1+\sqrt{1-a^{*2}})}\,.
\end{equation}

\begin{figure}[htp!] 
\centering
\includegraphics[width=0.45\textwidth]{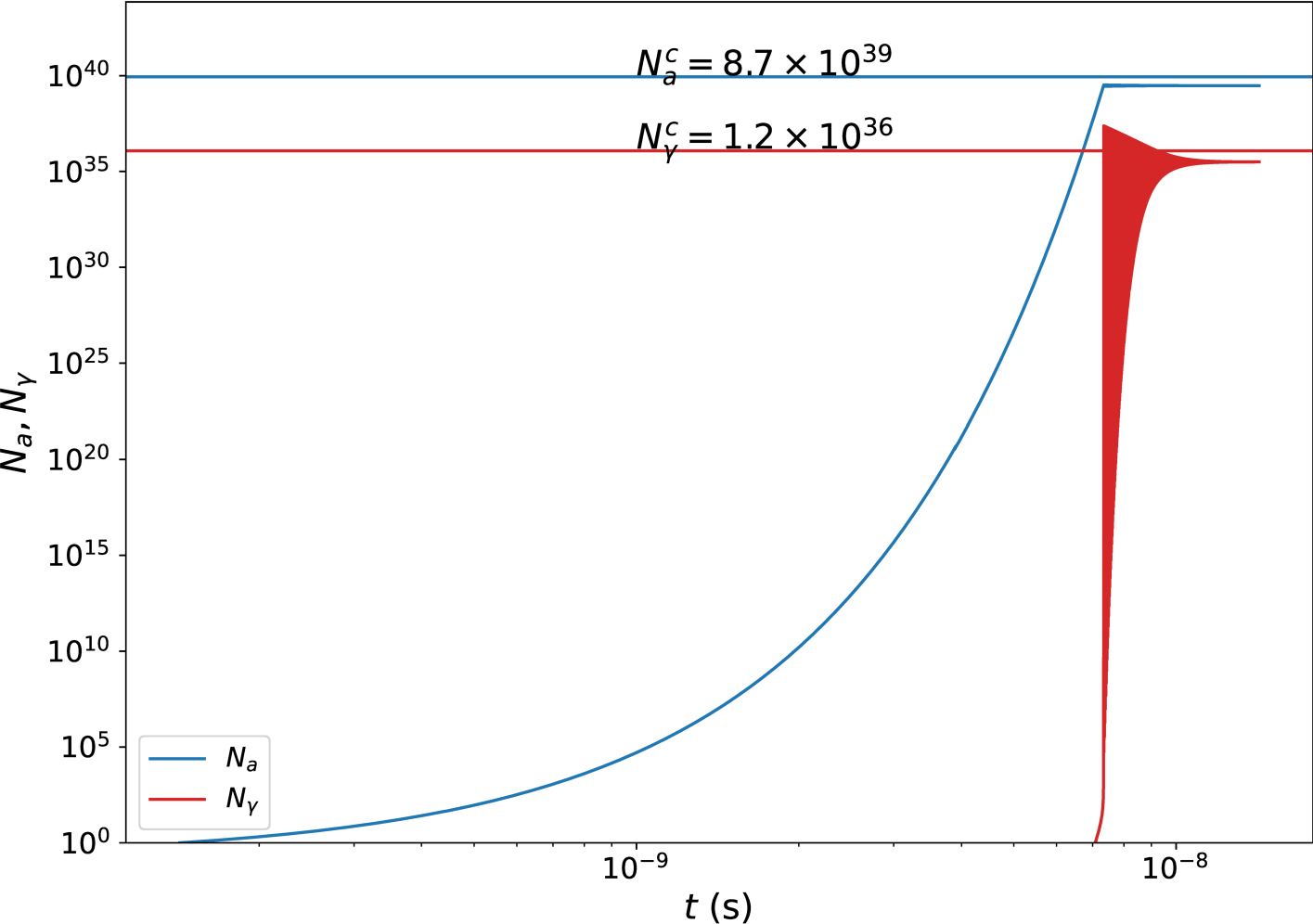}
\caption{ Numerical solutions for the Boltzmann system Eq.~\eqref{eq:SDOE_NAG} of axions with $\mu=1\mathrm{eV}$ inside the cloud due to the formation of instability around a BH of mass $M_{BH}=5\times 10^{-11}\mathrm{M}_\odot$ and spin $a^*=0.7$}\label{fig:NaNg}
\end{figure}

The previous results can be verified with a numerical solution of the Boltzmann system~\eqref{eq:SDOE_NAG}. An example is shown in Fig.~\ref{fig:NaNg} for the particular case of an axion mass $\mu=1\mathrm{eV}$ and a BH mass $M =3.35\times 10^{-11}\mathrm{M}_\odot$. We can clearly see the exponential growth of the particle numbers at early times, followed by an oscillatory asymptotic behavior at late times. These oscillations are around the asymptotic values $N_{\gamma}^c$ and $N_{a}^{c}$ previously predicted. 

\begin{figure}
\centering
\includegraphics[width=0.45\textwidth]{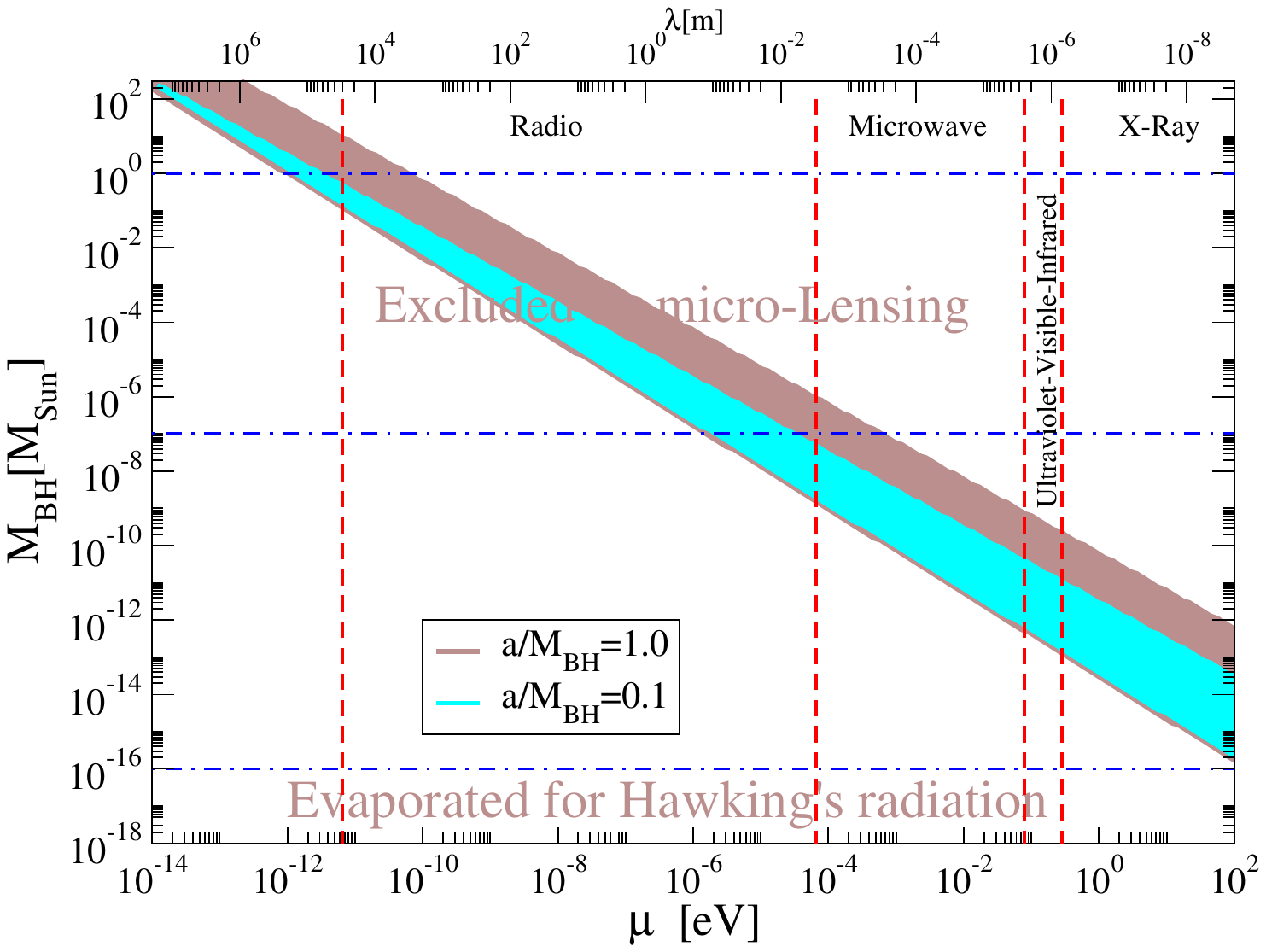}
\caption{Parameter space for an axion/ALP involving the minimal conditions for superradiance with emission of photons due to the coupling of axion/ALP particles with photons for two values of the black hole spin: $a^*=1$ (cyan region) and $a^*=0.1$ (blue region). The upper axis shows the wavelength of photons emitted by the axion/ALP cloud. For definiteness, $f_a=10^8$ GeV has been fixed.  
}
\label{fig:ALP_parameter_space}
\end{figure}

Now we can scan the region of interest in the $(\mu,M_{BH})$ parameter space for the adimensional spin parameter $a^*=a/M_{BH}$, so that there will be a possible emission of photons due to superradiance with axion/ALP properties. The most important condition is that the number of critical axions $N_a^c$ is reached at a time comparable with the age of the universe $\tau_H\approx10^{10}$ years, e.g.
\begin{equation}
t=\frac{\ln{N^c_a}}{\Gamma_s-\Gamma_a}<\tau_H\,.\label{eq:contidion_photons}
\end{equation}
Eq.~\eqref{eq:contidion_photons} imposes a restricted region in the parameter space $(\mu,M_{BH})$ that is shown in Fig.~\ref{fig:ALP_parameter_space} for two representative values of $a^*$, namely $a^*=1$ (cyan region) and $a^*=0.1$ (brown region). The black holes within those regions had time to develop an axion/ALP cloud that will emit photons because of the process described in \cite{Rosa:2018Sti}. The photons emitted would have energy $E_\gamma=\mu/2$. Wavelengths of such possible emitted photons are shown on the upper axis of Fig.~\ref{fig:ALP_parameter_space} that allows us to conclude that only black holes with masses $M_{BH}< M_\odot$ can be detectable in the radio wavelength if axion/ALP particles have masses bigger than $\mu>10^{-11}\mathrm{eV}$. Lower masses will produce photons with wavelengths too large to be detectable because they are below the CMB temperature. 

This will be the region where we will concentrate since it could be possible to have photon emission produced by PBH.

\section{Bounds on the PBH fraction from the intensity of cosmic background radiation} 
In the previous sections, we have shown that the formation of scalar field clouds around black holes is a feasible physical scenario, and we have delineated the parameter space under which an axion/ALP triggers superradiance, ultimately leading to photon emission from the surrounding environment of black holes of specific masses. Here we will compute the flux of photons emitted by PBHs and compare the predicted flux with current limits on the observation of the intensity of cosmic background diffuse radiation. In order to avoid an excess of photons, upper limits in the number of PBH can be obtained. The constraints are computed for four distinct scenarios arising from the combination of two PBH spatial distributions and two particle physics models. We assume a monochromatic mass distribution for the PBHs, which constitute a fraction $f_{PBH}$ of the total cold dark matter, such that $\Omega_{PBH} = f_{PBH} \Omega_{CDM}$. We further consider additional dark matter contributions from ultralight scalars. For the spatial distribution of PBHs, we consider:

\begin{enumerate}
    \item A homogeneous distribution of extragalactic PBHs.
    \item A scenario where PBHs follow hierarchical structure formation, treating galaxies as dark matter overdensities with a Navarro-Frenk-White (NFW) density profile.
\end{enumerate}

For each of these spatial distributions, we analyze two particle physics cases: (i) the QCD axion, where the decay constant $f_a$ is fixed by the particle mass i.e., \cite{ParticleDataGroup:2024cfk}
\begin{equation}
f_a=5.69\left(\frac{\mu}{\textrm{meV}}\right)^{-1}10^9\textrm{GeV} \label{fa_axion}
\end{equation}
and the second case corresponds to ALPs, where the decay constant $f_a$ and the mass $\mu$ are treated as independent variables.  The present limits on $f_a$ found by a diverse and complementary set of experiments \cite{VanTilburg:2015oza,Oswald:2021vtc,Filzinger:2023zrs,Zhang:2022ewz,Raffelt:1996wa, ADMX:2009iij, IAXO:2013len, d2021collider, Aiko:2023EW, barth:2013cast, aprile:2017xenon1t, Akerib:2017LUX, fu:2017PandaxII} yield a reasonable value of $f_a=10^{11}$ GeV. The limits obtained for ALP particles for $f_{BH}$ that we will show will consider this value.
Next we will review how to compute the photon flux for each scenario.
 
\label{sec:fluxes}
\subsection{Constraints from Extragalactic Background light}
\begin{figure*}[t]  %
  \centering
\includegraphics[width=\textwidth]{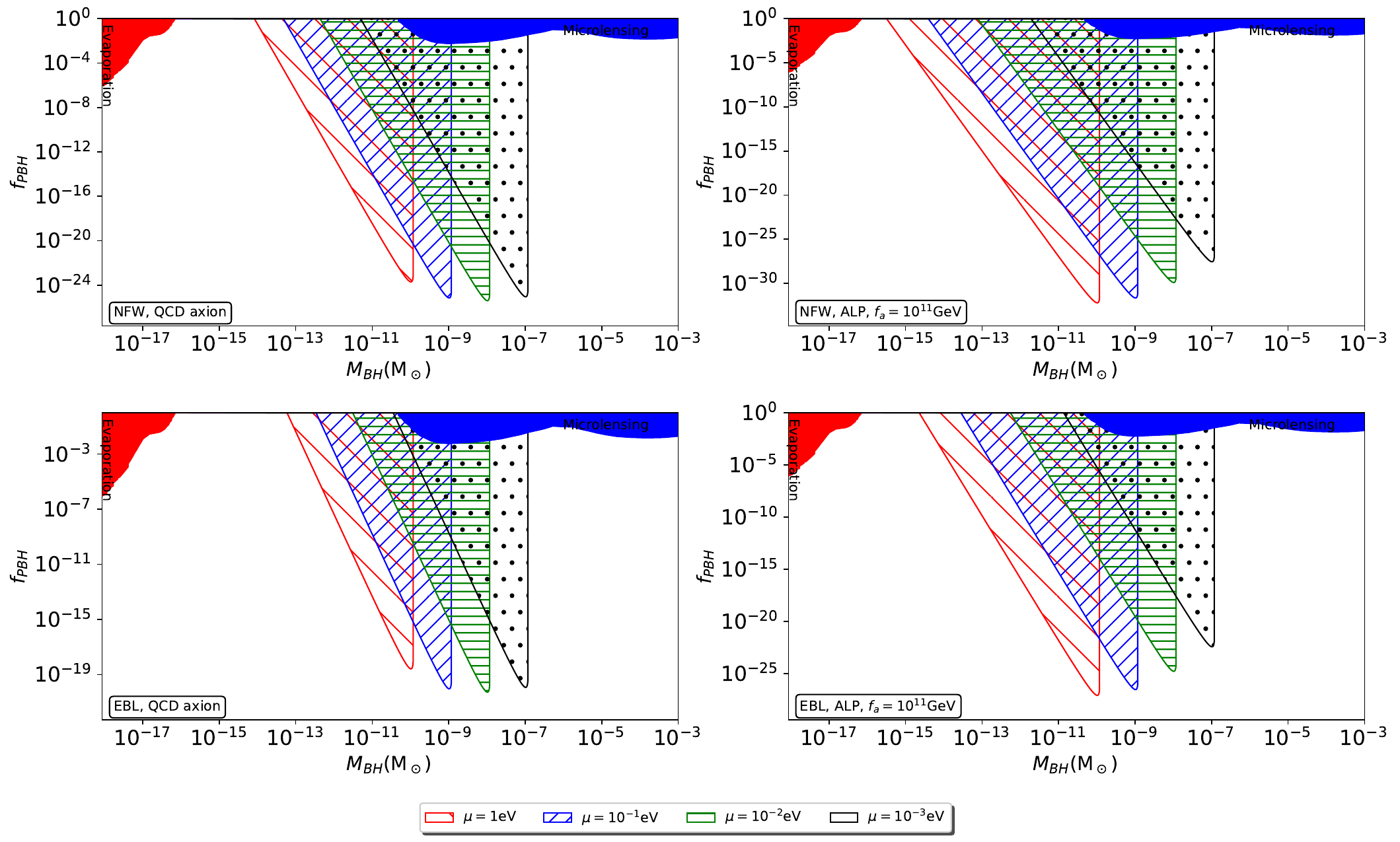}
  \caption{$f_{PBH}$ as a function of $M_{BH}$ in the window of interest for local and extragalactic spectrum}
  \label{fig:bounds}
\end{figure*}
In this section, we assume that all the light measured at a point is the result of integrating the whole population of PBHs distributed through space, assuming that there is no evaporation due to our mass choice (so in principle we assume that the whole population of PBH is conserved). Supposing that we are located at the center of a spherical shell of radius $a$, the integrated light from all PBHs with scalar clouds can be calculated as the integrated flux of photons per unit area per unit time distributed over a solid angle.
Following \cite{Overduin:2004dark}, the flux of energy per unit area reaching us from a single source is given by
\begin{equation}
dQ_P=\frac{R^{2}(t)L(t)}{4\pi R_0^4 r^2}\, ,
\end{equation}
where the scale factor at the time the light reaches us is given by $R(t_0)=R_0$, $r$ is the source coordinate distance, and $L(t)$ is the luminosity or the energy per unit time. The presence of two factors $(R(t)/R_0)$ reflects Hubble's energy and number effects. \\
Assuming that we are located in a spherical shell of radius $r$ and that the sources are located at an extended $dr$ distance, the volume of the shell can be written as 
\begin{equation}
 dV=4\pi R^2 r^2 c dt\, .
\end{equation}
 The total energy measured due to the presence of multiple sources (in our case of interest PBHs) inside the shell, can be calculated as the product of the total number of PBHs per their individual intensity \cite{Overduin:2004dark}
 \begin{equation}
 dQ=n_P dQ_P dV=c n(t)\tilde{R}L(t)dt\, ,
 \end{equation}
 where $n_P$ is the number density of PBH's and $L(t)$ is the luminosity. For the last term of the expression, we use the fact that, under the hypothesis of conservation of the density of PBH, it is possible to replace $n_P$ as a comoving number quantity $n(t)\equiv(R(t)/R_0)^3 n_P=\tilde{R}^3 n_P$. Due to the selected mass window, the total quantity of PBH remains the same at present $n\sim n_0$.
The luminosity emitted by a source between the range $\lambda$ and $\lambda+d\lambda$ as \cite{Overduin:2004dark}
\begin{equation}
    L=\int F(R,\lambda)d\lambda\sim F(R,\lambda)\lambda_0
\end{equation}
where $F(R,\lambda)=E_0\frac{dN}{d\lambda dt}$ is the Spectral Distribution Energy (SED). This is the integrated light from many galaxies, which has been emitted at various wavelengths and redshifted by various amounts, which is distributed in a waveband centered on $\lambda_{0}$ when it arrives to us. We refer to this as the spectral intensity of the EBL at $\lambda_{0}$.
The intensity of the shell  as observed at wavelength $\lambda_{0}$ is given by
\begin{equation}
4\pi dI_{\lambda}\equiv dQ_{\lambda,obs}=-cn(t)\tilde{R}^{2}(t)E_0 F[\tilde{R}(t)\lambda_{0},t]\lambda_0 dt,
\end{equation}
where the factor $4\pi$ converts from an all-sky intensity to one measured per steradian and $E_0$ is the energy of a single photon, $n(t)$ is the comoving number of galaxies and $c$ is the speed of light. 

We are interested in a monochromatic emission profile, where the central line depends on the mass of the axion decaying to two photons. 
In the scenario of interest, the relevant photons are those whose mean free path coincides with the radius of the black hole as $\Gamma_e N_\gamma^c$. Thus, if we use $N_\gamma^c=\frac{\Gamma_S}{A\Gamma_{a\gamma\gamma}}$, this implies $\Gamma_{a\gamma\gamma}N_a\rightarrow 2\Gamma_s N_{a}^c$. Therefore,
\begin{equation}
\frac{dN_\gamma}{dE dt}=2 N_a \Gamma_{s} \delta(E-E_{\gamma})\, ,
\label{eq:SED}
\end{equation}
where $E_{\gamma}=\mu/2$. 
In the scenario of the formation of $PBH's$, we consider a matter-dominated universe
\begin{equation}
n_{PBH}=\Omega_{PBH}\frac{\rho_c^0}{M_{BH}}\tilde{R}^{-3}\, ,
\end{equation}
with $\Omega_{PBH}=f_{PBH}\Omega_{CDM}$. $f_{PBH}$ is the fraction of DM that can be conceived as PBH, and $\rho_c^0$ is the critical density at $t_0$. Thus,
\begin{equation}
I=\frac{c}{4\pi}f_{\mathrm{PBH}}\Omega_{\mathrm{CDM}}\rho_c^0\lambda_{\gamma}^0 E_0\int_{t_{rec}}^{t^0}\frac{2N_a\Gamma_{s}\delta(\lambda-\lambda_{\gamma})}{\tilde{R}(t)M_{BH}(t)}dt\, .
\end{equation}
Changing the integration variable to $\lambda$ in a matter-dominant scenario $(R(t)\propto t^{2/3(1+w)}$ with $w_{MD}=0)$ and using
\begin{equation}
1+z=\frac{\lambda_0}{\lambda}=\frac{R_0}{R}=\left(\frac{t_0}{t}\right)^{2/3}\, ,
\end{equation}
we integrate over the Dirac's delta and obtain the total intensity. Deriving the result with respect to $\lambda_0$, we obtain the differential intensity distribution in the present time per unit wavelength (where we divide by $E_0$ to adjust the units \cite{Overduin:2004dark}). $N_a$ and $M_{BH}(t)$ are calculated in $t\rightarrow t(\lambda_\gamma^0)$
\begin{equation}
I_\lambda=\left|\frac{dI}{d\lambda_\gamma^0}\right|=\frac{9c}{8\pi}f_{\mathrm{PBH}}\Omega_{\mathrm{CDM}}\rho_c^0 t^0 \Gamma_{s}\left(\frac{\lambda_\gamma}{\lambda_\gamma^0}\right)^{3/2}\frac{N_a(t)}{M(t)}\frac{1}{\lambda_{\gamma^0}}\, .
\label{eq:dIdlambda}
\end{equation}
 
With~\eqref{eq:dIdlambda}, it is possible to find boundaries over the value of $f_{PBH}$ if the theoretical value exceeds the data for a certain region. We can set limits over the values of $f_{PBH}$ for a QCD-like axion and ALP for different masses using interpolation from data reported in \cite{Overduin:2004dark, cooray:2016extragalactic} and summarized in Table \ref{table:I_data} for our cases of interest.
 The bounds obtained by demanding $I_\lambda < I_{\lambda}^{Obs}$ are shown in Figure \ref{fig:bounds}. For the axion case the bounds are shown in the left-lower panel while the case for ALP with $f_a=10^{11}$ GeV in the right-lower panel.
\begin{table}[t]
\centering
\caption{Experimental measurements of the intensity of cosmic background radiation at some wavelengths relevant for our study. The values shown were obtained by interpolation of data reported in \cite{Overduin:2004dark,cooray:2016extragalactic}.}
\label{table:I_data}
\begin{tabularx}{\columnwidth}{c c X}
\hline\hline
$\mu$ (eV) & $\lambda$ &
$I_{\lambda}^{\mathrm{Obs}}$
($\text{\AA}\,\mathrm{s\,sr\,cm^{2}})^{-1}$ \\
\hline
$1$        & 98 nm (UV-C)              & 658   \\
$10^{-1}$  & 980 nm (IR-A)             & 235   \\
$10^{-2}$  & 9.86 $\mu$m (far IR)      & 1335  \\
$10^{-3}$  & 98.6 $\mu$m (far IR)      & 29719 \\
\hline\hline
\end{tabularx}
\end{table}

\subsection{Constraints on $f_{PBH}$ for clustered PBHs in the Milky-Way}

Now, we are interested in calculating the total luminosity associated with the number of photons yielded in the damped regime due to the cloud of axions of a galactic distribution of PBH. For simplicity we will consider that PBH form the DM halo and that the distribution follows a Navarro-Frenk-White profile. Taking the flux associated with a monochromatic distribution with energy per photon $E=\mu c^{2}$, we see that the luminosity per black hole is
\begin{equation}
    L=\Gamma_e N_\gamma\frac{\mu}{2}\, .
\end{equation}

The number of PBHs per unit area $(\tilde{N})$ can be obtained integrating over a density DM distribution in our galaxy, where we put an observer $O$ at a distance $d=8kpc$ from the galactic center, who observes a region around the galactic halo with an aperture of $40\mathrm{arcsec}$ from the galactic plane to the line of sight \cite{NFWparameters}. Using $\rho_s=0.4\mathrm{GeV cm^{-3}}$, $r_s=21\mathrm{kpc}$ for a NFW profile and dividing by $M_{PBH}$, we have all the parameters needed to calculate the number of PBH's inside the local vicinity
\begin{equation}
    \tilde{N}=\frac{1}{M_{\mathrm{PBH}}}\int_{\Delta\Omega}d\Omega\int_0^{\infty}\rho(r(l,\Omega))dl\,,
\end{equation}
where $\rho(r, l)$ is the distribution function of the DM in the local media integrated over the line of vision that along with the integral over the solid angle provides the total distribution of mass. In a general scenario where the photons depend explicitly on the axion decay, the flux can be calculated as
\begin{equation}
    \phi=\tilde{N}\left(\int_{E_1}^{E_2}\frac{dN}{dt dE}dE\right)\, ,
    \label{eq:flux}
\end{equation}
 The emission spectrum is proportional to the photon distribution, and, since we are considering a monochromatic spectrum, we can write again the SED per photon as \eqref{eq:SED}. 

To compare with the differential intensity data \cite{Overduin:2004dark, cooray:2016extragalactic} to obtain the maximal contribution of these possible candidates to CDM, we divide the flux \eqref{eq:flux} over $\Delta\Omega \lambda_\gamma$ to obtain the differential intensity. Adjusting the units to $\mathrm{CU's}=\mathrm{photons}~s^{-1}~cm^{-2}~\mathrm{str}^{-1}~\mathring {A}^{-1}$ we get
\begin{align}
    \left|\frac{dI}{d\lambda_{\gamma}^0}\right|=&2.18\times 10^{-3} f_{\mathrm{PBH}}\left(\frac{f_a}{10^{11}\mathrm{GeV}}\right)^2\left(\frac{M_{BH}}{10^{-13}M_\odot}\right)^6\nonumber\\
&\left(\frac{a^*}{0.9}\right)\left(\frac{\mu}{10^{-3}\mathrm{eV}}\right)^7\left(\frac{1}{C_{a\gamma\gamma}}\right)^2\mathrm{CU's}\, . \label{difintensityNFW}
\end{align}

Comparing the intensity calculated Eq. \eqref{difintensityNFW} with the observed intensity for a fixed mass of the axion/ALP particle shown in Table \ref{table:I_data} we set bounds on $f_{PBH}$ shown in the upper panel of Figure \ref{fig:bounds}. Two cases were considered: left panel for the case of an axion particle with $f_a$ given by Eq. \eqref{fa_axion} and for an ALP the bounds are shown in the top-right figure for $f_a=10^{11}$ GeV.


\section{Conclusions}\label{sec:Conclusions}

In this work, based on the results presented by Rosa and Kephart in \cite{Rosa:2018Sti}, we have found the region of parameters $(\mu, M_{BH},a^*)$ where emission of photons is possible (see Figure \ref{fig:ALP_parameter_space} on a time scale lower than the age of the Universe.  By using the current limits on the astrophysical flux of photons measured at Earth, it is possible to find reliable constraints on the fraction of PBH that might constitute the DM of the universe. Severe constraints of the order of $f_{BH}<10^{-17}$ were found. 

We can see that the bounds that can be obtained with ``shining black holes" are stronger than the previous bounds on the fraction of PBHs in the mass range $10^{-13} M_{\odot}<M_{BH}<10^{-7} M_\odot$ in all the scenarios we have studied and summarized in Figure \ref{fig:bounds}. This is because the knowledge of diffuse radiation at wavelengths where superradiance can trigger emission of photons in PBHs is well known. As usual, the bounds are model-dependent and in our case we have assumed the simplest hypothesis in order to show that this mechanism, if ultralight particles exist, can give relevant constraints. A more detailed analysis will be performed elsewhere.

\subsection*{Acknowledgements}
YG thanks CONACYT for financial support. This work was partially supported by Programa para el Desarrollo Profesional Docente; Direcci\'on de Apoyo a la Investigaci\'on y al Posgrado, Universidad de Guanajuato, research Grants No. 036/2020, 099/2020; CONACyT M\'exico under Grants No. A1-S-17899, No. 286897, No. 297771, No. 304001; and the Instituto Avanzado de Cosmolog\'ia Collaboration.



\end{document}